\documentclass[prb,twocolumn,showpacs,preprintnumbers,amsmath,amssymb,superscriptaddress]{revtex4}
\usepackage{graphicx}
\usepackage{dcolumn}
\usepackage{bm}

\usepackage{epstopdf}

\usepackage{color}

\begin{document}

\title{Multipolar, magnetic and vibrational lattice dynamics in the low temperature phase of uranium dioxide}

\author{R. Caciuffo}
\affiliation{European Commission, Joint Research Centre, Institute
for Transuranium Elements, Postfach 2340, D-76125 Karlsruhe, Germany}

\author{P. Santini}
\affiliation{Dipartimento di Fisica, Unit\`a CNISM di Parma, Universit\`a di Parma, via G.P. Usberti 7/a, 43100 Parma, Italy}

\author{S. Carretta}
\affiliation{Dipartimento di Fisica, Unit\`a CNISM di Parma, Universit\`a di Parma, via G.P. Usberti 7/a, 43100 Parma, Italy}

\author{G. Amoretti}
\affiliation{Dipartimento di Fisica, Unit\`a CNISM di Parma, Universit\`a di Parma, via G.P. Usberti 7/a, 43100 Parma, Italy}

\author{A. Hiess}
\affiliation{Institut Max von Laue-Paul Langevin, 6 rue Jules Horowitz, BP 156, F-38042 Grenoble Cedex 9, France}

\author{N. Magnani}
\affiliation{Lawrence Berkeley National Laboratory, 1 Cyclotron Road Berkeley, CA  94720-8175 USA}

\author{L.-P. Regnault}
\affiliation{ISPSMS-MDN, UMR-E 9001, CEA/UJF-Grenoble 1, INAC, F-38054 Grenoble, France}

\author{G. H. Lander}
\affiliation{European Commission, Joint Research Centre, Institute
for Transuranium Elements, Postfach 2340, D-76125 Karlsruhe, Germany}

\date{\today}

\begin{abstract}
We report the results of inelastic neutron scattering experiments performed with triple-axis spectrometers to investigate the low-temperature collective dynamics in the ordered phase of uranium dioxide. The results are in excellent agreement with the predictions  of mean-field RPA calculations emphasizing the importance of multipolar superexchange interactions.
By comparing neutron scattering intensities in different polarization channels and at equivalent points in different Brillouin zones, we show the mixed magneto-vibrational-quadrupolar character of the observed excitations. The high energy resolution afforded by the cold triple-axis spectrometer allowed us to study in detail the magnon-phonon interaction giving rise to avoided crossings along the $[00\xi]$ reciprocal space direction.
\end{abstract}


\maketitle

\section{Introduction}
Systems with active orbital degrees of freedom display peculiar dynamical properties.  These include well-known single-ion phenomena like the dynamical Jahn-Teller or the quadrupolar Kondo effect, but also  propagating excitations more complex than conventional magnetic spin-waves or excitons, and involving quadrupoles or higher-rank multipoles alone or in combination with dipoles. The identification and modeling of such collective excitations may shed light onto the scarcely known world of multipolar two-ion interactions, whose nature and whose form usually remain elusive \cite{santini09}. Indeed, static properties (e.g., the character of multipolar order or the response to macroscopic external perturbations)  are typically affected by only a few of the many possible microscopic couplings between active multipoles, and moreover only space-integrated information on these couplings can be extracted (e.g., the value of the mean-field constant).
When quadrupoles are order parameters in a phase transition, electronic and phononic contributions to two-ion quadrupolar interactions cannot usually be unambiguously separated. As a consequence, in $d$ or $f$ systems with quadrupolar order the relative importance of phonons  ("\textit{Jahn-Teller}") and exchange ("\textit{electronic}") mechanisms in the phase transition has been debated for years and cannot be inferred from static properties.
Conversely, in the dynamics many active multipoles and the detailed microscopic interactions among them play a role, and much more information can be extracted by modeling experimental data on multipolar dynamics, in much the same way as the detailed form of spin exchange couplings can be extracted by modeling spin-wave spectra.

Uranium dioxide is an ideal compound to study the effects of multipolar interactions on the collective lattice dynamics. The first inelastic neutron scattering (INS) investigations of antiferromagnetic (AF) UO$_{2}$ were performed in the late $1960$s \cite{dolling65,dolling66,cowley68} and gave convincing evidence of a strong interaction between magnon and phonon excitations below the N\'{e}el temperature, T$_{N}$ = 30.8 K. However, interpretations of the results in terms of simple Heisenberg-type spin-wave theories were unsatisfactory. Later INS studies performed without \cite{buyers85} and with neutron polarization
analysis \cite{caciuffo99,amoretti99,blackburn05,caciuffo07} provided a
detailed picture of the low-energy quasi-particle dynamics in UO$_{2}$, revealed a large anisotropy of the dipolar exchange two-ion coupling, and confirmed the presence of features not compatible with a description including only vibrational and magnetic exchange interactions. Indeed, further experimental evidence suggested that quadrupolar interactions play an important role in stabilizing the ground state of UO$_{2}$, as a static Jahn-Teller distortion of the oxygen cage \cite{faber75} and long-range order of electric quadrupoles \cite{wilkins06} occur along with the first-order magnetic transition at T$_{N}$.
The fact that the magnetic order is of the 3-$\textbf{k}$ type \cite{burlet86} indicates that quadrupolar interactions have both magneto-elastic (ME) and purely electronic two-ion contributions, produced by the same mechanism responsible for the anisotropy of the dipolar exchange, namely superexchange (SE) interactions in presence of orbital degeneracy \cite{santini09}. With ME coupling alone, in fact, the 1-$\textbf{k}$ structure would be more stable than the 3-$\textbf{k}$ one \cite{giannozzi87}, which is stabilized by the SE coupling \cite{mironov03}. These quadrupolar interactions have a significant impact also on the dynamics of UO$_{2}$.
\begin{figure*} 
\includegraphics[width=12.0cm]{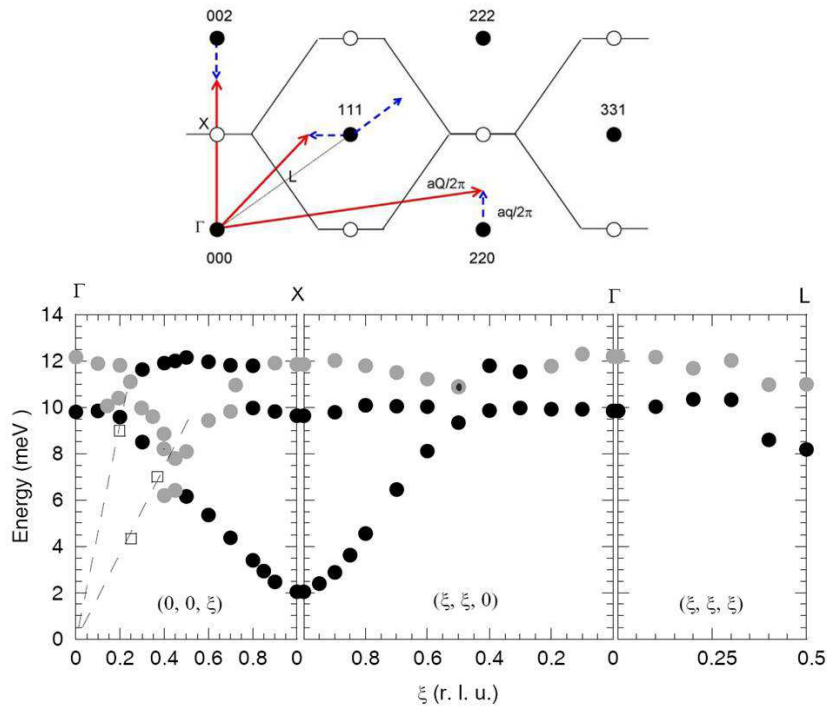}
\caption{(Color online). The dispersion of the mixed magneto-vibrational-quadrupolar modes in the low-temperature phase of UO$_{2}$ is shown along the principal crystallographic directions of the cubic lattice \cite{caciuffo99,buyers85}. Neutron groups have been collected in the constant-Q mode, at the reciprocal lattice points $(0,0,2- \xi)$,$(1- \xi,1- \xi ,1)$, and $(1 + \xi,1 + \xi,1 + \xi)$.
Grey symbols indicate qualitatively smaller intensity than the black symbols. The broken lines and open squares indicate acoustic phonon branches measured at 270 K, \cite{caciuffo99}. Excitations around 12 meV correspond to an optical quadrupolar
branch. The experiments provide evidence of mixing between magnons and phonons, in the forms of large  avoided crossings (ACs), for instance at $\xi \sim$ 0.45 along the $[0 0 \xi]$ direction. A portion of the UO$_{2}$ $(1\overline{1}0)$ reciprocal lattice plane is shown in the upper panel. Filled and empty circles are points corresponding to crystal and magnetic structure Bragg reflections, respectively. Red (solid) arrows indicate neutron momentum transfer vectors; blue (dashed) arrows indicate quasi-momentums of collective excitations.}
\label{dispersion}
\end{figure*}

The magnetic low-temperature dispersion curves provided by previous investigations using various triple-axis spectrometers at different neutron sources are shown in Fig. \ref{dispersion}. The experimental data contain evidence of magnon-phonon interactions at several position in reciprocal space, the most evident one being the avoided crossing (AC) between the magnon and the transverse acoustic (TA) phonon branches propagating in the $[0 0 \xi]$ direction at $\xi$ = 0.45 ($\xi = aq/2\pi$, $a$ being the lattice parameter of the cubic unit cell). Along the same $[0 0 \xi]$ direction, two distinct optical branches and additional excitations occur in the energy range between 9 and 12 meV. These features are not compatible with a scenario assuming 3-$\textbf{k}$ magnetic order and dipole-only interactions. Indeed, the 3-$\textbf{k}$ magnetic structure of UO$_{2}$ has $Pa\overline{3}$ symmetry and the magnetic modes along the $\Gamma$-$X$
direction should belong to one acoustic branch and a single doubly degenerate optical branch (see Fig.~\ref{maps001}(a)).
The experimental picture is not reproduced in a dipole-dipole scheme even if all the SE anisotropic coupling parameters allowed in the distorted $Pa\overline{3}$ structure are taken into account \cite{santini09}. Giannozzi and Erd\"{o}s tried to solve the puzzle by attributing the optical branch at about 12 meV to quadrupole excitations \cite{giannozzi87}. However, assuming a special form of phonon-mediated quadrupole-quadrupole coupling, they obtained calculated intensities for the quadrupolar modes smaller than the experimental values by two order of magnitudes.

A satisfactory model was eventually provided by Carretta et al. in ref. [\onlinecite{carretta10}], where they report the outcome of extensive theoretical calculations that take into account SE magnetic-dipole and electric-quadrupole coupling, together with free phonons and ME interactions.
These calculations show that quadrupolar interactions are an essential ingredient to understand the dynamics of UO$_{2}$ in its magneto-quadrupolar ordered phase. The optical branch at $\sim 12$ meV corresponds
to propagating quadrupolar fluctuations carrying along a magnetic component. These excitations can be visualized as
wave-like oscillations of the two locally-orthorhombic quadrupoles ($\mathcal{Q}_{xy}$, $\mathcal{Q}_{x^2-y^2}$, where the $z$-axis is along the local moment direction) accompanied by the precession of a small spin component \cite{carretta10}. The model also predicts additional excitations in the energy range between $\sim$ 8 and $\sim$ 12 meV\cite{buyers85}, and the existence of a dispersive low-energy acoustical quadrupolar branch in the $[0 0 \xi]$ direction, characterized by a very weak INS intensity.

In this paper, we present the results of triple-axis INS experiments performed with thermal neutrons and polarization analysis, and with unpolarized cold neutrons. The main goals of these experiments were to find evidence of the weak intensity modes predicted in
Ref.~[\onlinecite{carretta10}], including the acoustic quadrupole-wave branch, and to better characterize the AC between magnons and TA phonons at $(0,~0,~0.45)$. The results we obtained are in excellent agreement with the theoretical predictions. By comparing neutron scattering intensities in different polarization channels and at equivalent points in different Brillouin zones, we show the mixed magneto-vibrational character of the observed excitations. The high energy resolution afforded by the cold triple-axis spectrometer allowed us to clearly separate the two modes participating to the AC at $(0,~0,~0.45)$ and to follow the change from magnon-like to phonon-like character of the two branches.

The remainder of this paper is structured as follows. In Sec. \ref{theory} we briefly summarize the theoretical model developed in Ref.~[\onlinecite{carretta10}]; experimental details are given in Sec. \ref{exp det}, and the results obtained are reported and discussed in Sec. \ref{res dis}. A short summary and conclusions are found in Sec. \ref{concl}.

\section{Theoretical outline} \label{theory}
In this work, collective excitations in UO$_{2}$ are calculated in the mean field random phase approximation \cite{jensen91,carretta10} (MF-RPA). The dynamics are driven by time-dependent effective fields transmitted by phonons and two-ion electronic exchange. Fields that are dipolar in character, such as those associated with magnetic exchange, couple to the total angular momentum $\textbf{J}$ of the U ions and give rise to conventional magnon modes. Higher-rank multipolar interactions transmit fields that couple with higher-rank multipoles and may drive multipolar waves.
For quadrupoles or even-rank multipoles these effective fields are physically equivalent to a time-dependent crystal-field. Fields transmitted by exchange are nearly instantaneous on the timescale of the collective excitations, whereas fields transmitted by phonons (necessarily even-rank) may display important retardation effects.
The response of each ion to the fluctuating fields is encoded in the set of energy-dependent single-ion susceptibilities $\chi_{A B} (\omega)$ ($A$ and $B$ denote dipoles or multipoles), which are calculated from self-consistent MF eigenstates.
Poles in $\chi_{A B} (\omega)$ are located at the MF energy gaps and describe single-ion excitations, which become dispersive excitation branches in the crystal. We label a branch as multipolar if the matrix element between the corresponding MF levels is zero for dipole operators, and non-zero for higher-rank operators.

\begin{figure}       
\includegraphics[width=8.0cm]{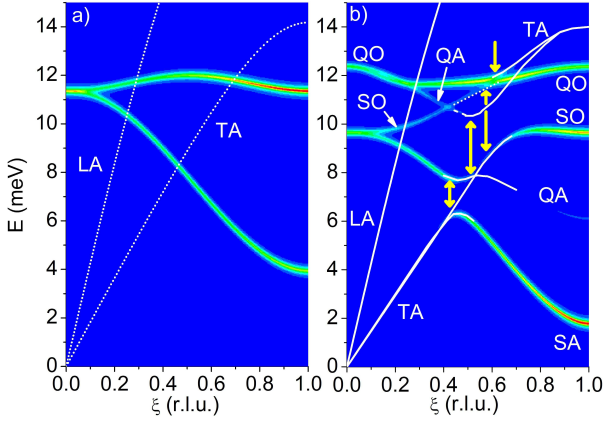}
\caption{(Color online). Dipolar magnetic INS cross section along $[0 0 \xi]$, calculated with the model described in Ref.~[\onlinecite{carretta10}] and discussed in this work. The color scale represents intensity in arbitrary units. In panel a) the intensity map shows the results of RPA spin-wave calculations neglecting both superexchange and magnetoelastic quadrupole-quadrupole  interactions. Bare acoustic phonon dispersion curves are shown by white lines. Panel b) shows the intensity map obtained by taking into account quadrupole-quadrupole couplings. White lines show modes with sizeable phonon component.
Labels specify the character of the excitation branches: SA=spin acoustic, SO=spin optical, QA=quadrupolar acoustic, QO=quadrupolar optical, TA = transverse acoustic phonon, LA = longitudinal acoustic phonon. Vertical double arrows indicate avoided crossing positions.}
\label{maps001}
\end{figure}

\begin{figure}       
\includegraphics[width=7.5cm]{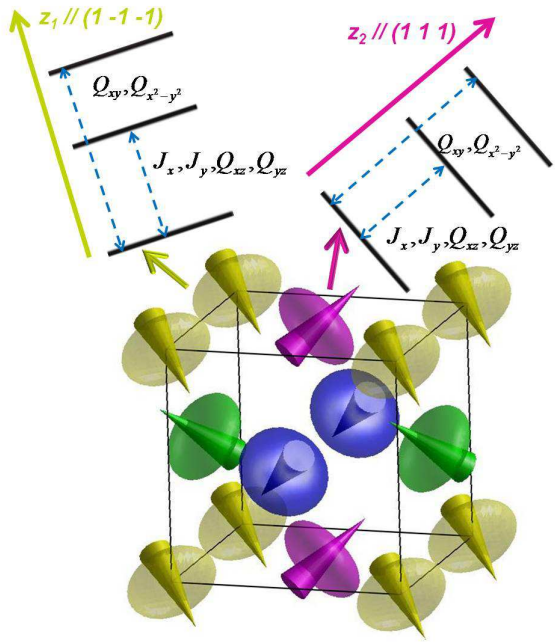}
\caption{(Color online). Arrangement of the ordered magnetic moments (cones) and
quadrupoles (ellipsoids) in the 3-${\bf k}$ phase. The 4 sublattices are labeled by different colors.
The figure also illustrates the splitting of the ground $\Gamma_5$
triplet below $T_N$ for two sublattices. The labels of non-vanishing off-diagonal matrix elements are referred to the local $z$-axis.
\label{rpa} }
\end{figure}

The U ions in UO$_{2}$ are tetravalent, with a $5f^{2}$ electronic configuration. The crystal field (CF) ground state is a $\Gamma_{5}$ triplet \cite{magnani05}, carrying magnetic dipoles ($J_{x}$, $J_{y}$, $J_{z}$),  $\Gamma_{5}$ electric quadrupoles $\mathcal{Q}_{\alpha}$ transforming as $xy$, $yz$, and $zx$, and a doublet of $\Gamma_{3}$ electric quadrupoles transforming as
$3z^{2}-r^{2}$ and $x^{2}-y^{2}$. In the ordered 3-$\textbf{k}$ phase, the $\Gamma_{5}$ triplet is split into three singlets, labeled by the $z$ components of an effective $S$ = 1 pseudo-spin operator ($M = 0, \pm 1$). In a reference frame with the $z$-axis along the local magnetic moment direction, that is one of the equivalent $\langle 1 1 1\rangle$ directions, the operators $J_{z}$ and $\mathcal{Q}_{3z^{2}-r^{2}}$ are diagonal and, therefore, static (see Fig.~\ref{rpa}). The remaining six operators are non-diagonal and fluctuating: the first MF transition, between $M$ = -1 and $M$ = 0, is associated with fluctuations of $J_{x}$,  $J_{y}$, $\mathcal{Q}_{zx}$, and $\mathcal{Q}_{yz}$. This mixed magnetic-quadrupolar mode yields the conventional spin-waves of Fig.~\ref{maps001}(a) if quadrupolar interactions are neglected. However, the two quadrupolar operators provide coupling of this excitation to phonons and this is the origin, for example, of the anticrossings around 7 meV near  $\xi =0.45$ (TA-SA mixing, see Fig.~\ref{maps001}(b)and Fig.~\ref{disp00z}) and around 11 meV near $\xi =0.6$ (TA-SO mixing, see Fig.~\ref{maps001}(b)).
The second MF excitation from $M$ = -1 to $M$ = 1 is purely quadrupolar at the MF level being associated with fluctuations of $\mathcal{Q}_{x^{2}-y^{2}}$ and $\mathcal{Q}_{xy}$. The dispersion of collective excitations stemming from this MF mode is due entirely to two-ion quadrupolar interactions. The dynamics are again further complicated by the coupling with phonons provided by the electric quadrupoles and manifested as ACs between electronic and vibrational modes.

As the ordered lattice of UO$_{2}$ contains four independent U sites\cite{santini09}, a total of eight branches originate from the two single-ion MF excitations. Four of them are acoustic branches, two spin (SA) and two quadrupolar (QA), and four are optical branches, again two spin (SO) and two quadrupolar (QO) branches. The number of acoustic phonon branches is twelve, when their folding into the smaller Brillouin zone of the ordered phase is considered. 

The Hamiltonian to be considered, $H=H_{SS}+H_{\mathcal{QQ}}+H_{P}+H_{ME}$, is the sum of terms corresponding to SE dipole-dipole ($H_{SS}$)
and quadrupole-quadrupole ($H_{\mathcal{QQ}}$) interactions, free phonons ($H_{P}$), and magnetoelastic phonon-quadrupole couplings ($H_{ME}$).
The number of free parameters defining $H$ is reduced to a minimum by limiting SE coupling to nearest-neighbors, and including only the quadrupoles of $\Gamma_5$  symmetry that order below $T_N$. For instance, for the pair of U ions at $0$~$0$~$0$ and $1/2$~$1/2$~$0$, we write

\begin{eqnarray}
H_{SS} &=& J [S_{z}(1) S_{z}(2)+ \delta (S_{x}(1) S_{x}(2)+S_y(1) S_y(2))] \nonumber \\
H_{\mathcal{QQ}} &=& K^{SE} [\mathcal{Q}^S_{xy}(1) \mathcal{Q}^S_{xy}(2) + \nonumber \\
 && \delta (\mathcal{Q}^S_{yz}(1) \mathcal{Q}^S_{yz}(2)+\mathcal{Q}^S_{xz}(1) \mathcal{Q}^S_{xz}(2))],
 \label{H}
\end{eqnarray}

keeping only three of the six symmetry allowed SE parameters, and assuming equal anisotropy $\delta$ for dipolar and quadrupolar SE interactions. $S_{\alpha}$ are components of the effective spin operator $S$ = 1, and $\mathcal{Q}^{S}_{\alpha} \propto \mathcal{Q}_{\alpha}$ are quadrupolar operator equivalents. For instance, $\mathcal{Q}^{S}_{xy} = S_{x}S_{y}+S_{y}S_{x}$.
The Hamiltonian for the remaining eleven bonds is obtained by rotating Eq.~(\ref{H}).

The phonon term $H_{P}$ is calculated within the rigid-ion model of Ref.~[\onlinecite{dolling65}], and in $H_{ME}$
only phonon-induced modulation of the oxygen cage are considered:
\begin{equation}
H_{ME}=\sum_{{\mathbf R}} \sum_{\alpha=xy,xz,yz} Q^S_{\alpha}({\mathbf R})[g_A \gamma_{A,\alpha}({\mathbf R}) + g_B \gamma_{B,\alpha}({\mathbf R})],
 \label{HME}
\end{equation}
\noindent
where $\gamma_{A,\alpha}({\mathbf R})$ and $\gamma_{B,\alpha}({\mathbf R})$ ($\alpha=xy, xz, yz$) are the $\alpha$ component of the two $\Gamma_5$
normal modes ($A$, $B$) of the oxygen cage around the U ion at ${\mathbf R}$ \cite{inoue64}. The five free parameters ($J$, $K^{SE}$, $\delta$, $g_A$, and $g_B$) are determined by comparison with experimental results.

Dynamical susceptibilities are calculated in RPA by considering fluctuations
around the 4-sublattice MF configuration, explicitly including spins, quadrupoles and phonons \cite{carretta10}. The RPA system is solved numerically and the inelastic neutron cross section is then obtained from the absorptive part of the dynamical magnetic susceptibility \cite{santini06}.
Within this approximation, phonons act as transmitting medium for effective two-ion quadrupolar couplings which can
be described in terms of a long-range retarded interaction $K^{ph}_{\alpha,\alpha^\prime}({\mathbf R}, {\mathbf R}^\prime ; \omega)$ between the quadrupoles $Q^S_{\alpha}(\mathbf R)$ and $Q^S_{\alpha^\prime}(\mathbf R^\prime)$. The physical picture is simple and recalls the mechanism by which conduction electrons transmit the RKKY interaction: the quadrupole
$Q^S_{\alpha^\prime}(\mathbf R)$ by Eq.(\ref{HME}) induces a distortion ($\gamma_{A/B \alpha}({\mathbf R})$) of the oxygen cage centered at $(\mathbf R)$. This distortion
corresponds to a superposition of lattice phonons and hence propagates with finite speed through $H_P$ to a different site $(\mathbf R^\prime)$.
If the resulting displacement pattern of the oxygen cage at $(\mathbf R^\prime)$ has nonzero projection onto the local normal modes ($\gamma_{A/B \alpha^\prime}({\mathbf R^\prime})$), it induces a polarization of $Q^S_{\alpha^\prime}(\mathbf R^\prime)$. This is equivalent to an effective quadrupolar retarded interaction: by writing ${\mathbf R}={\mathbf l}+{\mathbf h}$, where ${\mathbf l}$ spans
the Bravais lattice of the ordered phase and ${\mathbf h}$ spans the four sublattices, and by Fourier-transforming on ${\mathbf l}$, we find
\begin{eqnarray}
&&K^{ph}_{\alpha,\alpha^\prime}({\mathbf q}; {\mathbf h}, {\mathbf h}^\prime ; \omega)= e^{i ({\mathbf q}-{\mathbf G})\cdot ({\mathbf h}^\prime - {\mathbf h})}
\sum_{{\mathbf G},\sigma}
[g_A F^A_{\alpha \sigma}(-{\mathbf q}+{\mathbf G}) \nonumber \\
&&+g_B F^B_{\alpha \sigma}(-{\mathbf q}+{\mathbf G})]
[g_A F^A_{\alpha^\prime \sigma}({\mathbf q}-{\mathbf G})+g_B F^B_{\alpha^\prime \sigma}({\mathbf q}-{\mathbf G})]\nonumber \\
&&\times D_\sigma({\mathbf q}-{\mathbf G};\omega),
 \label{KEFF}
\end{eqnarray}
where
$F_{\alpha \sigma}({\mathbf q})= \hbar\Theta_{\alpha \sigma}({\mathbf q})/\sqrt {8m_O \epsilon_\sigma({\mathbf q})}$.
${\mathbf G}$ spans zero and the three basis vectors of the reciprocal lattice of the ordered phase and $\sigma$
labels the nine phonon branches. $D_\sigma({\mathbf q};\omega)$ is the free phonon propagator and $\Theta$ is the
projection of the $\alpha$ component of the local normal mode on the displacement pattern of phonon ${\mathbf q} \sigma$. Finally,
$m_O$ is the oxygen mass and $\epsilon_\sigma({\mathbf q})$ are the bare phonon energies.

Fig.~\ref{maps001}(a) shows the dipolar magnetic INS cross section calculated along $[0 0 \xi]$ by taking into account the most general-symmetry allowed expression of magnetic anisotropic exchange, but neglecting two-ion quadrupole-quadrupole coupling in the RPA system. The SA and the SO branches are degenerate at the $\Gamma$ point, and the SO branch has the same energy gap at $\Gamma$ and $X$.
In this calculation magnetic and vibrational modes are not coupled and no ACs are observed. Note that in the ordered phase the zone boundary corresponds to $\xi$ = 0.5 whereas the paramagnetic zone boundary $\xi$ = 1 ($X$ point) is a zone-center point equivalent to $\xi$ = 0 ($\Gamma$ point). Hence, all excitation energies are symmetric around $\xi$ = 0.5 and a second, specular SA branch also exists. However, the INS cross section is not symmetric around $\xi$ = 0.5 because of the unit cell structure factor which weights the four sublattices with different phases.

The dynamics change drastically if superexchange and magnetoelastic quadrupolar interactions are included. This is shown in Fig.~\ref{maps001}(b), where the intensity map corresponds to the set of parameters $J$ = 1.55 meV, $K^{SE}$ = 0.95 meV, $\delta$ = 0.25, $g_{A}$ = 6.7 meV/{\AA}, and $g_{B}$ = 42.9 meV/{\AA}. Many more excitations emerge in the INS cross section. Actually, at low $T$ there are as many as 20 elementary excitations for each wavevector : eight having a dominant electronic component and 12 having a dominant phonon component. Yet, only a few of them are actually detected by magnetic-dipole INS because of tiny dipolar matrix element (e.g., the QA branch) or because of tiny or vanishing INS structure factor.
The dynamics show a complex interplay of all these modes and the excitations usually display a mixed character. In particular, branches marked by white lines in Fig.~\ref{maps001}(b) have a sizeable vibrational component. In presence of quadrupolar coupling, the longitudinal acoustic (LA) phonon branch remains purely vibrational along the $[001]$ direction, whereas the two transverse acoustic (TA) branches acquire mixed spin-phonon character. The mixing is maximal near AC positions, approximately indicated by yellow arrows. The AC at  $\sim$7.2 meV and $\xi \sim$ 0.45 involves one of the two TA and the SA branches. It is due to the $\mathcal{Q}_{yz}$ and $\mathcal{Q}_{zx}$ quadrupoles, coupling phonons with the $\Delta M$ = 1 MF transition. The same quadrupoles are responsible for a second AC around 11 meV and $\xi$ = 0.6, mixing TA and SO modes. Further ACs occur at $\xi \sim$ 0.55 around 9 meV, and at $\xi \sim$ 0.6 around 12 meV. Both of them involve the $\Delta M$ = 2 MF transition and are due to coupling promoted by $\mathcal{Q}_{xy}$ and $\mathcal{Q}_{x^{2}-y^{2}}$ quadrupoles. The former mixes TA and QA modes, the latter TA and QO branches.

We stress that Fig.\ref{maps001} shows the {\it dipolar} magnetic INS cross section and hence the quadrupolar $\Delta M$ = 2 modes
emerge only where they acquire a magnetic component due to the mixing with $\Delta M$ = 1 spin-wave branches. This mixing
is produced by two-ion quadrupolar interactions, which also set the dispersion of quadrupolar waves. Hence, these waves are a very direct and accurate probe of the form of these otherwise elusive interactions. The mechanism producing the mixing can be visualized in the following way: if on a given ion (e.g., the one on the yellow sublattice in Fig.\ref{rpa})  $\mathcal{Q}_{xy}$ and $\mathcal{Q}_{x^{2}-y^{2}}$  fluctuate,
for the presence of two-ion quadrupolar interactions a different ion (e.g., the one on the purple sublattice in Fig.\ref{rpa}) feels a time-dependent quadrupolar molecular field, equivalent to a time-dependent distortion.
In the local (purple) $z$-axis, tilted with respect to that of the yellow site, this time-dependent field will induce oscillations of the local $\mathcal{Q}_{xy}$ and $\mathcal{Q}_{x^{2}-y^{2}}$ quadrupoles but also (off-resonantly) of the local $\mathcal{Q}_{xz}$ and $\mathcal{Q}_{yz}$ quadrupoles. The latter are locked to $\mathcal{J}_{x}$ and $\mathcal{J}_{y}$ and hence a magnetic oscillation is also induced on the purple site.
The result is a certain amount of  $\Delta M$ = 1 component in the $\Delta M$ = 2 branches, which turns out to be large for the QO branches, and small or nearly vanishing for the QA branch.

Of course, quadrupolar modes can also be indirectly detected at large $Q$ through the vibrational INS cross section where they display sizeable mixing with phonons (see below).
The vibrational spectra of UO$_{2}$ at low temperature are clearly complicated. We compare the observed excitations with a model that has only five parameters, so a precise fit should not be expected in every case. Nevertheless, as shown in the following figures the agreements are very good.

\begin{figure}       
\includegraphics[width=7.5cm]{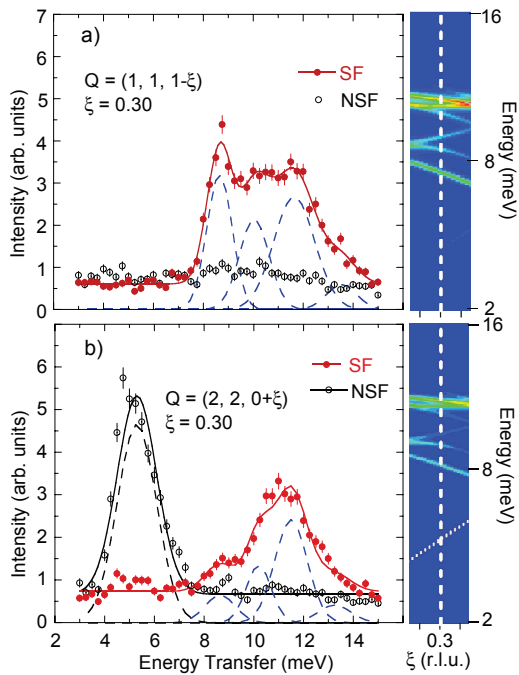}
\caption{(Color online). Constant-$\textbf{Q}$ energy scan recorded with IN22 at (a) $\textbf{Q} = (1, 1, 1)-(0, 0, 0.3)$, and (b)  $\textbf{Q} = (2, 2, 0)+(0, 0, 0.3)$, ($\textbf{P} || \textbf{Q}$; T = 2 K). Red (filled) circles represent neutron counts in the spin-flip (SF) channel, open circles are non-spin-flip (NSF) counts. The SF data are fitted to the sum of four gaussian line-shapes shown by dashed (blue) lines.The peak at 5 meV in the NSF channel at $\textbf{Q} = (2, 2, 3)$ is a TA phonon (black dashed line). The corresponding portion of the calculated intensity map is shown
as a guide to the energies of the excitations. The dashed (white) line gives the $q$ value at which the energy scan has been performed.
\label{csi03} }
\end{figure}

\section{Experimental details} \label{exp det}
The experiments were performed on the same single crystal used previously \cite{caciuffo99,amoretti99,blackburn05,caciuffo07}. The sample, of $\sim$
9 cm$^3$ in volume, was cut from a melt-growth crystal
boule of depleted uranium dioxide and aligned with the $(1,\bar{1},0)$ vector
perpendicular to the scattering plane. Rocking curve measurements show
a homogeneous mosaic spread of about 0.4 degrees, with a small portion
of the sample misoriented by $7$ degrees.

Measurements were carried out at the Institut Laue-Langevin (ILL), in
Grenoble, France. The unpolarized INS data were collected on the IN14 triple-axis
cold-neutron spectrometer, using vertically focusing pyrolytic graphite (PG-$002$) crystal as monochromator and analyzer. The instrument was in the long-chair geometry, with open-60$^{\prime}$-60$^{\prime}$-open
collimation. A beryllium filter was inserted between the sample position and the analyzer's collimator. Data were collected in the constant-$Q$ mode, with $k_{f}$ fixed at
1.5 \AA$^{-1}$.

Polarization analysis measurements were carried out on the thermal neutron IN22
triple-axis spectrometer, operated in
the fixed-$\bm{\mathrm{k}}_{f}$ mode ($k_{f}$ = 2.662 \AA$^{-1}$),
with Cu$_2$MnAl Heusler crystals as monochromator and analyser. A
pyrolytic graphite filter was used for second harmonic rejection.
The CRYOPAD device \cite{brown98,tasset01} was used to control the
polarisation $\mathbf{P}_{i}$ of
the incident
neutron beam and to analyze the longitudinal components of $\mathbf{P}_{f}$, the polarisation of the scattered beam. A
flipping ratio $R$ = 10.5$\pm$0.5 was estimated from the peak intensity of the transverse acoustic phonon mode measured at $\textbf{Q} = (2, 2, 0.3)$.
A standard liquid He cryostat was used to maintain the sample at 2~K on both spectrometers.

In this work, the ${\bf x}$-axis of the reference frame is defined to be
parallel to the momentum transfer $\mathbf{Q}$ =
$\mathbf{k}_{i}-\mathbf{k}_{f}$, the ${\bf z}$ axis is perpendicular to the
scattering plane and ${\bf y}$ forms a right-handed system.

A general expression of the cross-section for polarized neutrons,
describing the final polarization $\mathbf{P}_{f}$ of a scattered
neutron beam as a function of the incident beam polarization $\mathbf{P}_{i}$,
was derived by Blume\cite{blume63} and Maleyev \cite{maleyev63}.
For ${\bf P}_{i}$ and ${\bf P}_{f}$
parallel to the scattering vector
$\mathbf{Q}$ (${\bf x}$-${\bf x}$ geometry), the magnetic signal contributes only to the spin-flip (SF) channel, that is the neutron spin is reversed upon
scattering. This does not occur in the case of vibrational scattering,
which appears only in the non-spin-flip (NSF) channel.

Note that in a general case multipolar waves might be detected directly by nondipolar INS. Magnetic neutron scattering is produced by fluctuations of time-odd multipole moments like magnetic dipoles or octupoles, and the multipolar contributions to the neutron cross-section may enable purely non-dipolar excitations (with vanishing dipole-moment matrix element between initial and final states) to be directly probed \cite{levy77,sablik85}. This cross-section is rather small but may be detectable, especially its octupolar component. However, in the specific case of UO$_{2}$ there are no fluctuations of magnetic multipoles associated with the  $\Delta M$ = 2 transition. Indeed, magnetic multipoles with non-vanishing matrix elements within the CF ground state have $\Gamma_{4}$ symmetry. Since $\Gamma_{5} \times \Gamma_{5}$ contains $\Gamma_{4}$ only once, all these multipoles are proportional to one another and map onto the three pseudospin operators $S_{x}$, $S_{y}$ and $S_{z}$. Hence the same selection rules valid for dipoles hold for high-rank $\Gamma_{4}$ multipoles and $\Delta M$ = 2 transitions are forbidden. Thus, as already stressed above, the dispersion of quadrupolar waves can be mapped by conventional dipolar or vibrational neutron scattering by exploiting the fact that when these $\Delta M$ = 2 excitations propagate they may acquire a magnetic or phonon component from nearby $\Delta M$ = 1 or phonon branches.

\section{Experimental results} \label{res dis}
The theoretical model described in Sec.~\ref{theory} predicts the presence of more than two branches propagating along $[00\xi]$, with appreciable magnetic INS intensity near the crystallographic zone center, $\Gamma$. The extra branches have a mixed spin-optical, quadrupolar-optical, and quadrupolar-acoustic character. Since for $\xi \simeq 0.3$ the phonon component is small, these branches are expected to be measurable only in the SF channel.
To verify this predictions, we have performed scans corresponding to $\xi$ = 0.3 with IN22 in the polarization analysis mode of operation. As shown in Fig.~\ref{csi03}a, reporting spectra at $\textbf{Q} = (1, 1, 0.7)$, intensity is indeed observed only in the SF channel, and the spectra can be fitted to the sum of four not-fully-resolved gaussian line-shapes. This is confirmed by the results shown in Fig.~\ref{csi03}b, where we report equivalent spectra measured under the same conditions away from $\textbf{G} = (2, 2, 0)$. The peak at 5~meV, corresponding to a TA phonon, is observed only in the NSF channel at $\textbf{Q} = (2, 2, 0.3)$, its intensity being negligible at $\textbf{Q} = (1, 1, 0.7)$ because of the unfavorable structure factor ratio of about 20 compared to $\textbf{Q} = (2, 2, 0.3)$.

\begin{figure}       
\includegraphics[width=7.5cm]{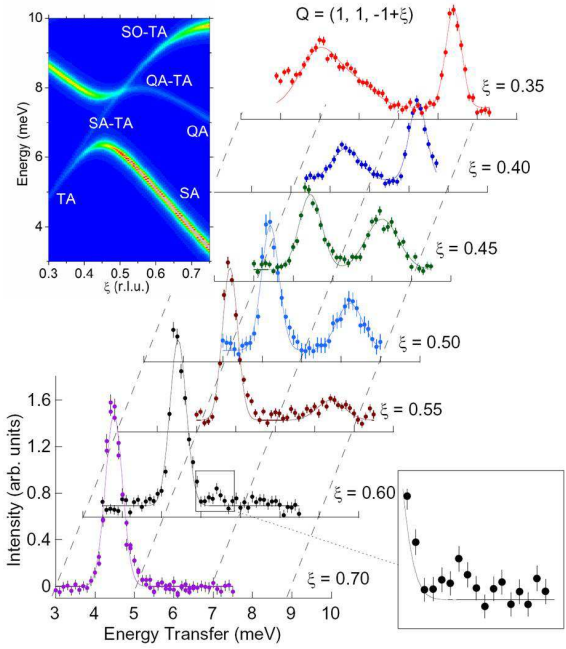}
\caption{Dispersion along the $[0 0 \xi]$ direction of the mixed magnon-phonon modes in UO$_{2}$ at 2 K, near the anticrossing position at $\textbf{q} = (0, 0, 0.45)$. The diagrams show detector counts normalized to monitor as a function of the energy transfer measured with IN14 at $\textbf{Q} = (1, 1, -1+\xi)$. A sloping background has been subtracted from the raw data. The inset shows calculated intensity maps in the energy-momentum transfer space (zoom of Fig.~\ref{maps001}(b)). For $\xi$ = 0.6, a small peak is visible at 6.5 meV (see Fig. \ref{qata}).
\label{disp00z} }
\end{figure}

One of the most striking feature in the low-energy collective mode spectra of UO$_{2}$  is the large avoided crossing between SA and TA modes at $\xi$ = 0.45 along $[0 0 \xi]$. Evidence of magnon-phonon interactions at this position was clearly given in the earliest investigation  \cite{dolling66}, but the energy resolution of previous studies was not sufficient to separate the interacting modes. Here, using the IN14 cold-neutron triple-axis spectrometer, we have been able to resolve the two excitations involved all along the AC region.

Fig.~\ref{disp00z} shows a sequence of constant-$\textbf{Q}$ energy scans collected at T = 2 K. The spectra have been measured in the constant-$\textbf{Q}$ mode at reciprocal space positions $\textbf{Q} = \textbf{G}+\textbf{q}$, with $\textbf{G} = (1, 1, -1)$ and $\textbf{q} = (0,0,\xi)$. Close to the magnetic zone center, the crystallographic $X$ point at $\xi$ = 1, only the SA excitation is observed, but the two interacting modes are clearly visible in proximity of the AC position. For $\xi > $~0.45
the phonon-like excitation is on the high-energy side, whereas it occurs on the low-energy side for $\xi <$~0.45. The two modes are clearly distinguished by the difference in width, which is due to the fact that their dispersion curves have a slope of opposite sign, corresponding to focusing and defocusing configurations.

\begin{figure}       
\includegraphics[width=7.5cm]{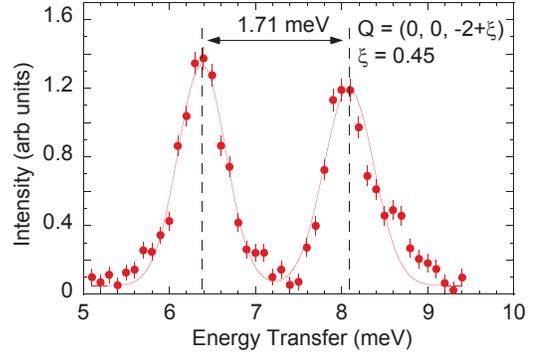}
\caption{(color online). Longitudinal constant-$\textbf{Q}$ energy scan at the anticrossing position $\textbf{q} = (0, 0, 0.45)$. The experimental data have been measured with the IN14 spectrometer at
$\textbf{Q} = (0, 0, -1.55)$, with the sample kept at T = 2 K. A sloping background has been subtracted from the raw data. The two Gaussian curves (solid, red lines) have equal width.
\label{ac00z} }
\end{figure}

The results of a longitudinal scan at $\textbf{Q} = (0, 0, -2)+(0, 0, 0.45)$ are shown in
Fig.~\ref{ac00z}. In this spectrometer configuration, a purely vibrational TA mode would have zero cross-section because the phonon polarization is perpendicular to $\textbf{Q}$. The fact that two peaks with comparable intensity are observed indicates that both excitations have 50$\%$ magnon-like and 50$\%$ phonon-like character. As stated above, the magnon-phonon coupling is here mediated by $\mathcal{Q}_{yz}$ and $\mathcal{Q}_{zx}$ quadrupoles, having non-zero matrix elements between the $M$ = -1 and $M$ = 0 MF single-ion states.

The origin of the excitations away from the AC can be probed by comparing spectra measured at equivalent positions in different Brillouin zones, and making use of polarization analysis. Magnetic and vibrational excitations are well separated, the
first appearing in the SF channel and the latter in the NSF one.
Moreover, phonon intensities increases with the square of the momentum transfer $Q$, whereas magnetic intensities decreases with increasing $Q$ following the square of the magnetic form factor $f(Q)$. However, near an AC
wavevector, peaks are visible in both SF and NSF channels due to their mixed magnetovibrational character.

\begin{figure*}       
\includegraphics[angle=270,width=12cm]{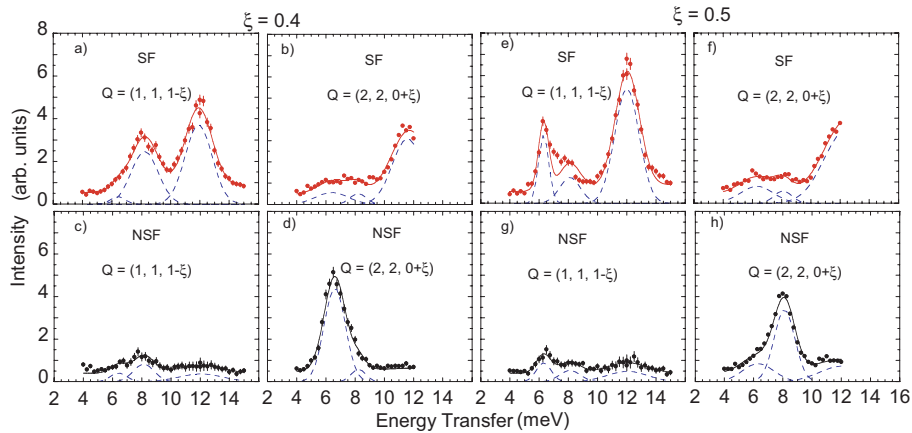}
\caption{(Color online). Constant-$\textbf{Q}$ energy scans recorded with IN22
at (a-d) $\textbf{Q} = (1, 1, 0.6)$ and $\textbf{Q} = (2, 2, 0.4)$; (e-h) $\textbf{Q} = (1, 1, 0.5)$ and $\textbf{Q} = (2, 2, 0.5)$, corresponding to the quasi-particle wave vector $\textbf{q} = (0, 0, 0.4)$ and $\textbf{q} = (0, 0, 0.5)$, respectively. Experimental data are fitted to the sum of three gaussian line-shapes, shown by dashed (blue) lines, and a constant background.
\label{IN220405}}
\end{figure*}

Figs. \ref{IN220405} show transverse constant-$\textbf{Q}$ energy scans taken with IN22 at equivalent
positions, away from $\textbf{G} = (1, 1, 1)$ and $\textbf{G} = (2, 2, 0)$ reciprocal lattice points. Data have been measured with the sample kept at 2~K in the ${\bf x}$-${\bf x}$ ($\textbf{P} || \textbf{Q}$) polarization geometry. The spectra in Fig.~\ref{IN220405}(a-d) corresponds to the quasi-particle wave vector $\textbf{q} = (0, 0, 0.40)$, those in Fig.~\ref{IN220405}(e-h) to $\textbf{q} = (0, 0, 0.50)$.

The spectra can be fitted to the sum of three Gaussians.
The peak at about 12 meV
in the SF channel corresponds to the spin precession accompanying the wave-like oscillations of the $\mathcal{Q}_{xy}$ and $\mathcal{Q}_{x^{2}-y^{2}}$ quadrupoles. This excitation would have negligible intensity in the absence of two-ion quadrupolar interactions. The two peaks at lower energy correspond to the same excitations shown in Fig. \ref{disp00z}. In the fit, their energy has been fixed to the values observed with better resolution on IN14.

At $\xi$ = 0.4, the peak at 6.4 meV is strong in intensity only in the NSF channel at $\textbf{Q} = (2, 2, 0.4)$, and is therefore of mainly vibrational character. On the other hand, the peak at 8.2 meV has the largest intensity in the SF channel at $\textbf{Q} = (1, 1, 0.6)$, and is therefore of dominant magnetic origin. Nevertheless, the mixed character of the excitations is revealed by the observation in both polarization channels of intensities larger than leakage corresponding to a flipping ratio $R$ of about 10.
At $\xi$ = 0.5, the situation is reversed. The peak at 6.2 meV is now strongest in the SF channel
at $\textbf{Q} = (1, 1, 0.5)$, and is mainly magnetic, whereas the peak at 8.2 meV has the largest intensity in the NSF channel at $\textbf{Q} = (2, 2, 0.5)$, and is therefore mainly vibrational in origin.

\begin{figure}       
\includegraphics[width=7.0cm]{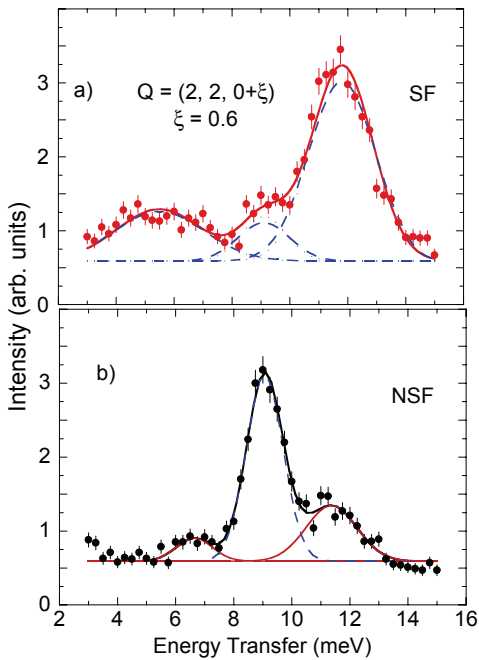}
\caption{(Color online). Constant-$\textbf{Q}$ energy scan recorded with IN22 at $\textbf{Q} = (2, 2, 0)+(0, 0, 0.6)$ ($\textbf{P} || \textbf{Q}$; T = 2 K). a) Neutron counts in the spin-flip (SF) channel; individual components of the spectrum are shown by dashed lines. b) Non-spin-flip (NSF) neutron counts. The central peak at $\sim$ 9 meV (blue dashed line) is a transverse acoustic phonon, whereas the weaker peaks at $\sim$ 6.6 meV and $\sim$ 11.4 meV (red full lines) correspond to mixed phonon-quadrupolar acoustic modes.
\label{qata} }
\end{figure}

\begin{figure}       
\includegraphics[width=7.5cm]{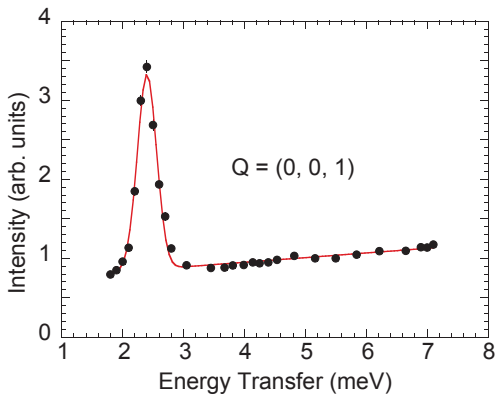}
\caption{(Color online). Constant-$\textbf{Q}$ energy scan recorded with IN14 at $\textbf{Q} = (0, 0, 1)$, corresponding to the X point of the crystallographic cell. The experimental data are fitted to the sum of a gaussian line-shape and a sloping background. The experimental error on the intensity is smaller than the size of the symbols.
\label{xpoint} }
\end{figure}

The theoretical model described in section \ref{theory} predicts the existence of a QA branch that should become nearly pure for $\xi>$  0.7. This mode can only be observed in the INS spectra by mixing with either phonon or magnetic branches.
Fig.~\ref{qata} shows the results of IN22 measurements at $\textbf{Q} = (2, 2, 0.6)$, carried out to verify this possibility. In the SF channel we observe three peaks, corresponding to the spin-acoustic, spin-optical, and quadrupolar-optical branches, in increasing order of energy. The SO branch is here mixed with TA phonons due to the large AC marked by the arrow in Fig.~\ref{maps001}. Also the NSF spectrum shows three peaks and, as discussed in the following Section, the peak at about 6.6 meV corresponds to a mixed QA-TA branch.
We emphasize that the intensity observed at 6.6 meV in the NSF channel is a factor 10 larger than the leakage from the SF channel expected for a flipping ratio $R$ of about 10.5. The weak feature appearing at about 6.5 meV in the equivalent scan recorded with IN14 at $\textbf{Q} = (1, 1, -0.4)$ (see box in Fig. \ref{disp00z}) provides further confirmation of this finding. On the other hand, a scan with high counts statistics performed at the magnetic zone centre, $\textbf{Q} = (0, 0, 1)$, only show the SA branch at about 2.4 meV, with no evidence of further excitations between 3 and 7 meV (Fig.~\ref{xpoint}).

\section{Discussion}
\begin{figure}       
\includegraphics[width=8.5cm]{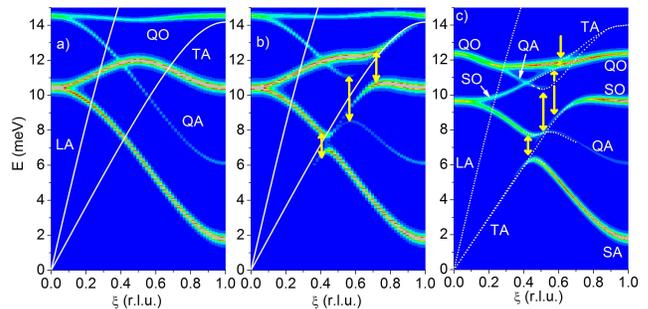}
\caption{(Color online). (a): Same calculation as in Fig.~\ref{maps001}b with the ME coupling removed, i.e., by setting $g_A=g_B=0$ in Eq.[\ref{HME}]. (b): Same calculation as in Fig.~\ref{maps001}b with high-energy phonons removed. This is done by artificially inserting a smooth cutoff
in $\Theta_{\alpha \sigma}({\mathbf q})$ in Eq.~(\ref{KEFF}) for phonons with energy $\epsilon_\sigma({\mathbf q})> 25 meV$. Panel c) is a reproduction of Fig. \ref{maps001}b.
The color maps have been re-scaled with respect to Fig.~\ref{maps001}b to enhance weak features.
\label{noJT} }
\end{figure}

 As Fig.~\ref{maps001}b well illustrates, the dynamics of UO$_{2}$ cannot be rationalized in terms of a simple paradigm, being the result of an intricate interplay of several sorts of motions. While all terms in Eqs.(\ref{H}), (\ref{HME}) contribute in a complex nonlinear way in determining the actual composition and dispersion of elementary excitations, a few general qualitative remarks can be made. The first one is that ACs involving phonons are a direct manifestation of the magnetoelastic coupling Eq.(\ref{HME}), as these would clearly not exist in the lack of such coupling.
For instance, Fig.~\ref{noJT}(a) shows how the INS cross section in Fig.~\ref{maps001}b changes when the ME coupling is removed, by keeping all remaining parameters fixed. Phonon and electronic modes are separated, and in the latter the "bare" SA, SO, QA and QO branches are clearly recognizable.
Even in the lack of ME coupling, however, the quadrupolar branches get a magnetic component by the two-ion quadrupolar superexchange.
The alternative calculation with quadrupolar SE removed while keeping the ME coupling is more problematic. In fact, SE has the leading role in stabilizing the ordered phase as it accounts almost entirely for the quadrupolar MF constant\cite{carretta10}. Hence, removing quadrupolar SE in the dynamics by leaving all other parameters unchanged seriously spoils the self-consistency of the calculation.

The theoretical model predicts several features that where not clearly observed in previous experiments and that are confirmed by the results presented here. In particular, besides confirming the QO character of the highest-energy branch propagating in the $[0 0 \xi]$ direction, the experiments provide evidence for the presence of a QA branch. Comparing Fig.~\ref{maps001} and Fig.~\ref{IN220405}, it appears that the peak observed in the SF channel at $\xi$ = 0.3 and 12 meV energy transfer
corresponds to the high-$E$ part of the QA branch.

In between $\xi \sim$ 0.3 and $\xi \sim$ 0.7 the QA branch experiences a very large AC with TA phonons to reemerge as a nearly pure QA mode
for $\xi \agt$ 0.7. The expected magnetic intensity is extremely weak, and in fact this
part of the branch is hardly visible in the map shown in Fig.~\ref{maps001} and it is not observed at $\textbf{Q} = (0, 0, 1)$ (Fig.~\ref{xpoint}).
On the other hand, more favorable conditions are met inside the AC range, where the QA mode is expected to mix with one of the TA phonon branches. In particular, $\xi \sim$ 0.6 is a good possible choice as the mode is fairly isolated in energy. Indeed,
in Fig.~\ref{qata}b the central peak at $\sim$ 9 meV (blue dashed line) is due to the TA-phonon component of the mixed TA-SO mode, the peak at $\sim$ 11.4 meV corresponds to the phonon component of the three mixed branches marked by white dashed lines in Fig.~\ref{maps001}, and the weaker peak at $\sim$ 6.6 meV corresponds to the mixed QA-TA branch. At the measured position, $\textbf{q} = (0, 0, 0.6)$, this branch has a sizeable vibrational component, and can therefore be detected in the NSF polarization channel. The peak at 6.6 meV provides therefore indirect evidence for the QA mode, as its intensity would be zero if the QA branch did not exist. The measured energy of the mode is slightly smaller than predicted ($\sim 7.7$ meV).
A discrepancy of this order is entirely reasonable given the simplified parametrization assumed in the theoretical model. On the other hand, the QA peak is not observed at $\xi$ = 1 (Fig.~\ref{xpoint}), because the mixing with the magnetic branch at this position is too weak.

An interesting and somewhat counterintuitive result of our simulations is the key role played by high-$E$ optical phonons in the
low-$E$ dynamics. One might expect only minor effects in view of the large energy mismatch between these phonons and the bare spin and quadrupolar
branches. However, the large modulation of the oxygen cage provided by the displacement pattern of some optical vibrations
(i.e., a large $\Theta_{\alpha \sigma}({\mathbf q})$ in Eq.~(\ref{KEFF})) enables important second-order mixing effects in spite of the
large energy differences. For instance, we show in Fig.~\ref{noJT}(b) a calculation where high-$E$ phonons have been artificially removed by leaving all parameters unchanged. The spectrum is very different from that of Fig.~\ref{maps001}b and is dominated by first-order mixing between
SA, QA and SO branches on one side, and TA phonon branches on the other side, leading to three clearly recognizable ACs. If optical phonons are included these ACs evolve into the three largest ACs marked in Fig.~\ref{maps001}b. Note how the second-order mixing with these phonons pushes down in energy all branches, and particularly the QO.
The result is that the mode intersects the TA phonons and produces the 4th small AC in Fig.~\ref{maps001}b.
Both the center and the width of the SA-TA AC at $\xi\sim 0.45$ are affected by the second-order mixing.

\section{Conclusions} \label{concl}

In this work, we report the results of extensive investigations on the collective quasi-particle dynamics in antiferromagnetic UO$_{2}$. Inelastic neutron scattering (INS) measurements have been performed with unpolarized-cold-neutron and polarized-thermal-neutron triple-axis spectrometers at the ILL. Exploiting the high energy resolution offered by cold neutrons, we have characterized magnon-phonon interactions in much greater detail than in previous investigations. The magnon-phonon coupling is particularly strong along the $[00\xi]$ direction at $\xi$ = 0.45. The avoided crossing involves the spin acoustic and one of the transverse acoustic phonon branches. Polarization analysis allowed us to shed light on the mixed character of the observed excitations. The experimental results have been discussed against the predictions of a mean-field random-phase approximation theory that provides convincing explanations for the several anomalous features in the UO$_{2}$ low-temperature dynamics that have resisted interpretation for almost half a century.

Our results demonstrate the existence of an optical branch corresponding to propagating quadrupolar waves, and provide evidence on a mixed vibrational-quadrupolar-acoustic branch. We also confirm the importance of superexchange multipolar interactions in determining the dynamics of systems with unquenched orbital degrees of freedom. Quadrupolar waves are not easily detected and have so far remained an elusive entity. In our experiments, the uranium atomic spins are caught in the act of surfing on quadrupolar waves that can therefore be detected by INS through the perturbation induced in the spin system. In $f$-electron materials, orbital degrees of freedom play a crucial role in a number of exotic phenomena, ranging from unconventional superconductivity to heavy-fermion behavior. The observation of multipolar waves is a key finding for the science and technology of this important class of materials.

\begin{acknowledgments}
We thank the ILL for the award of beamtime on the IN14, IN22, and IN3 spectrometers.
\end{acknowledgments}

\bibliography{uo2}
\end{document}